%% file: main.tex
\documentclass[letterpaper,twocolumn,10pt]{article}
\usepackage{usenix}

\input{Macros}

\begin{document}

\date{}

\title{Broken Gates: Re-evaluating Web Bot Defenses in the Age of LLM Agents}

\author{
{\rm Behzad Ousat, Nikita Turkmen, Lalchandra Rampersaud, Dillan Bailey, Selcuk Uluagac, and Amin Kharraz}\\
\{bousat, nturk005, lramp004, dbail042, suluagac, mkharraz\}@fiu.edu \\ \\
Florida International University
}

\maketitle

\thispagestyle{plain}
\pagestyle{plain}

\input{01-abstract}
\input{02-introduction}

\input{03-background}

\input{04-methodology}
\input{05-experiments}

\input{06-discussion}
\input{07-limitations}
\input{09-conclusion}

\bibliographystyle{plain}
\bibliography{refs}

\clearpage
\appendix
\input{10-appendix}

\end{document}

%% file: Macros.tex
\usepackage{tikz}
\usepackage{amsmath}

\usepackage{epsfig}
\usepackage{endnotes}
\usepackage{tikz}
\usepackage{fontawesome}
\usepackage{amsmath}
\usepackage{graphicx}
\graphicspath{ {./images/} }
\usepackage{caption}
\usepackage{subcaption}
\usepackage{listings}
\usepackage{fontawesome}
\usepackage{msc}
\usepackage{svg}
\usepackage{comment}
\usepackage{xcolor}
\usepackage{xspace}
\usepackage{booktabs}
\usepackage{colortbl}
\usepackage{multirow}
\usepackage{graphicx}
\usepackage{lscape}
\usepackage{rotating}
\usepackage{float}
\usepackage{url}
\usepackage{seqsplit}
\usepackage{hyperref}



\usepackage{listings}
\usepackage{color}
\definecolor{lightgray}{rgb}{.9,.9,.9}
\definecolor{darkgray}{rgb}{.4,.4,.4}
\definecolor{purple}{rgb}{0.65, 0.12, 0.82}
\usepackage{tikz}
\usetikzlibrary{arrows.meta, positioning, fit, calc}

\usepackage{tcolorbox}
\tcbuselibrary{listingsutf8}
\usepackage{listings}
\usepackage{xcolor}
\usepackage{fontenc}

\usepackage{amssymb}   

\usepackage{multirow}
\usepackage{amssymb}
\usepackage{pifont} 
\usepackage{makecell}

\definecolor{promptbg}{rgb}{0.97,0.97,0.97}
\definecolor{promptborder}{rgb}{0.2,0.6,0.86}

\newtcblisting{promptbox}[2][]{ 
  colback=promptbg,
  colframe=promptborder,
  listing only,
  listing options={
    basicstyle=\ttfamily\small,
    breaklines=true,
    columns=fullflexible,
    language=,
    escapeinside=||,
  },
  left=0.5mm,  
  right=0.5mm,
  top=0.5mm,
  bottom=0.5mm,
  boxrule=0.8pt,
  arc=2pt,
  enhanced,
  title={#2},  
  #1
}

\newcommand{\circled}[1]{%
\tikz[baseline=(char.base)]{
\node[shape=circle, draw, fill=black, text=white, scale=0.7, inner sep=2pt] (char) {#1};}}

\lstdefinelanguage{JavaScript}{
  keywords={typeof, new, true, false, catch, function, return, null, catch, switch, var, if, in, while, do, else, case, break},
  keywordstyle=\color{blue}\bfseries,
  ndkeywords={class, export, boolean, throw, implements, import, this},
  ndkeywordstyle=\color{darkgray}\bfseries,
  identifierstyle=\color{black},
  sensitive=false,
  comment=[l]{//},
  morecomment=[s]{/*}{*/},
  commentstyle=\color{purple}\ttfamily,
  stringstyle=\color{red}\ttfamily,
  morestring=[b]',
  morestring=[b]"
}

\definecolor{bypass}{RGB}{220,245,220}
\definecolor{fullpass}{RGB}{140,200,140}

\definecolor{codegreen}{rgb}{0,0.6,0}
\definecolor{codegray}{rgb}{0.5,0.5,0.5}
\definecolor{codepurple}{rgb}{0.58,0,0.82}
\definecolor{backcolour}{rgb}{0.95,0.95,0.92}

\newcommand{\ph}[1]{{\color{red}$wip$} }

\definecolor{darkpurple}{RGB}{85,26,139}

\lstdefinestyle{mystyle}{
    backgroundcolor=\color{backcolour},   
    commentstyle=\color{codegreen},
    keywordstyle=\color{magenta},
    numberstyle=\tiny\color{codegray},
    stringstyle=\color{codepurple},
    basicstyle=\ttfamily\footnotesize,
    breakatwhitespace=false,         
    breaklines=true,                 
    captionpos=b,                    
    keepspaces=true,                 
    numbers=left,                    
    numbersep=5pt,                  
    showspaces=false,                
    showstringspaces=false,
    showtabs=false,                  
    tabsize=2
}
\def\BibTeX{{\rm B\kern-.05em{\sc i\kern-.025em b}\kern-.08em
    T\kern-.1667em\lower.7ex\hbox{E}\kern-.125emX}}

%% file: 01-abstract.tex
\begin{abstract}

The rapid emergence of web automation across commercial crawlers, solver services, and LLM-based browser agents has created new challenges for defenses that must distinguish legitimate automated activity from scalable abuse. In response, modern bot defenses increasingly replace explicit challenges with risk-based decisions over browser, session, identity, and behavioral signals.
This paper explores how current automation stacks can bypass these defenses end-to-end, evaluating seven commercial solver services and eight LLM-based browser agents against four widely deployed defenses. We find that solver services largely defeat challenge-based defenses with a 100\% success rate, but collapse to an average 23\% success rate on reCaptcha v3. Browser agents encounter a different bottleneck: the browser and account state presented to the verifier. Controlled experiments across browser configurations show that an established consumer account and persistent Chrome state strongly affect reCaptcha~v2/v3 acceptance in our deployments, whereas Cloudflare Turnstile exhibits substantially greater sensitivity to the agent's browser control path. Reconstructing the established browser environment enables previously rejected agents to pass without model retraining.
Once established, this trusted environment transfers across the agents and defenses in our testbed, adding no measurable defense-specific marginal cost to the surrounding workflow. These results expose an inherent security tension in modern non-interactive defenses: to eliminate friction, the signals they trust must be accessible to everyday human users, yet this accessibility inherently guarantees that adversaries can acquire, reproduce, and transfer the same identity signals to scale automated abuse.
\end{abstract}

%% file: 02-introduction.tex
\section{Introduction}

\noindent
Web automation has become a routine part of the modern web ecosystem.
Commercial crawlers, scripted scanners, AI assistants, and LLM-based browser agents~\cite{yang2025agenticweb} can now navigate websites, complete forms, and execute multi-step workflows at scale.
While much of this automation is legitimate, the same capabilities also enable credential stuffing, large-scale scraping, fraud, vulnerability probing, and other forms of automated abuse.
Web services therefore rely on \emph{web bot-management} systems: defenses that inspect a client at an interaction point and decide whether to admit, challenge, or block it, with the goal of distinguishing legitimate users and benign automation from potentially malicious clients~\cite{google_recaptcha_about}.
These systems have traditionally made that decision through challenge-based Captchas.
Over time, they have shifted away from relying solely on explicit challenges. Providers increasingly combine, or replace, puzzle-solving with behavioral analysis, browser execution signals, reputation information, and risk-based scoring. Non-interactive defenses such as reCaptcha~v3~\cite{recaptcha_v3_about} and Cloudflare Turnstile~\cite{cloudflare_turnstile} exemplify this shift: rather than presenting a puzzle at a single decision point, they assess signals collected during page load and interaction, and only fall back to an explicit challenge, if at all, when that assessment is inconclusive. This move is driven largely by usability and conversion costs; friction at the challenge step measurably drives users away, and by the recognition that a one-time puzzle is a narrow checkpoint compared to signals gathered across an entire session.

This adaptation has sustained the competition between attackers and defenders, but it has also made the boundary these systems rely on less stable. Deployed defenses of both kinds still assume that automated actors remain behaviorally distinguishable from legitimate users, an assumption LLM-based browser agents directly challenge. These agents operate through ordinary web interfaces to navigate websites, interpret page content, submit forms, adapt to the page feedback, and carry multi-step workflows to completion from natural-language instructions alone. Automation is therefore no longer limited to brittle scripts that fail at hard interaction points. It can combine real-browser execution, feedback adaptation, and goal-directed task completion in ways that more closely resemble genuine human use.

A longer-standing weakness concerns the explicit challenge itself. Third-party solver services allow attackers to outsource Captcha challenges at low cost~\cite{gao2022demystifying,jin2023secure}, and credential stuffing, account takeover, and large-scale abuse campaigns routinely use these services as an operational step, allowing high-volume automation to continue despite deployed anti-bot controls~\cite{barkworth2022detecting,singh2024effective,li2021good}. 
Most prior work investigating the capabilities and limitations of these defense mechanisms has focused on exactly this step, solving the explicit challenge, whether through image recognition, audio transcription, or, more recently, LLM-based reasoning on puzzles~\cite{sivakorn2016m,ye2018yet,ieee11129490,deng2024oedipus,wang2026cognition}.

Reaching a protected action, however, is an end-to-end workflow: an automated client must arrive at the target page, present a browser environment and identity the defense is willing to trust, solve a challenge only if one is actually raised, and then carry the surrounding task through to completion. This distinction matters because the attack surface is moving. As deployments shift toward non-interactive defenses, there is frequently no puzzle to solve at all, and a view centered on challenge-solving misses where these defenses actually make their decision. Modern bot defenses have moved the security bottleneck from solving explicit challenges to reproducing the trusted execution context. When that context is transferable, reconstructing it can remove much of the defense's marginal cost from the automated attack workflow.


This observation leads to three questions illustrated in Figure \ref{fig:argument_map}. First, where does current automation fail in the end-to-end workflow? Second, when the defense raises no challenge, which property of the execution environment determines the verdict? Third, once that property is reconstructed, what cost does the defense continue to impose?

\begin{figure}[t]
\centering
\resizebox{0.9\columnwidth}{!}{%
\pgfdeclarelayer{background}
\pgfsetlayers{background,main}
\begin{tikzpicture}[
  font=\small,
  line cap=round, line join=round,
  every node/.style={align=center},
  title/.style={font=\bfseries\small},
  gloss/.style={font=\scriptsize, text=black!52},
  defn/.style={font=\scriptsize\itshape, text=black!42},
  ask/.style={font=\scriptsize, text=black!58},
  pill/.style={
    draw=none, fill=black, text=white, font=\bfseries\scriptsize,
    rounded corners=1.8pt, inner xsep=5.5pt, inner ysep=2.3pt
  },
  spine/.style={line width=0.9pt, black!62},
  dot/.style={circle, fill=black!62, inner sep=0pt, minimum size=2.4pt},
]
\def\xl{-2.2}
\def\xr{2.2}

\node[title] (root) at (0,0) {End-to-end bypass};
\draw[spine] (root.south) -- ++(0,-0.32) node[dot] (fork1) {};

\draw[spine] (fork1) -| (\xl,-0.88);
\draw[spine] (fork1) -| (\xr,-0.88);
\node[title] (inter) at (\xl,-1.08) {Interactive};
\node[gloss, below=0.6pt of inter] (interg) {challenge exists};
\node[defn, below=0.3pt of interg] (interd) {hCaptcha / reCaptcha~v2};
\node[title] (non) at (\xr,-1.08) {Non-interactive};
\node[gloss, below=0.6pt of non] (nong) {no explicit puzzle};
\node[defn, below=0.3pt of nong] (nond) {reCaptcha~v3 / Turnstile};

\draw[spine] (interd.south) -- ++(0,-0.3) -| (0,-2.58);
\draw[spine] (nond.south) -- ++(0,-0.3) -| (0,-2.58);
\node[dot] at (0,-2.58) {};
\node[pill] (rq1) at (0,-2.92) {RQ1};
\node[ask, below=2.2pt of rq1] (rq1q) {Where is the bottleneck?};
\draw[spine] (rq1q.south) -- ++(0,-0.3) node[dot] (fork2) {};

\draw[spine] (fork2) -| (\xl,-4.05);
\draw[spine] (fork2) -| (\xr,-4.05);
\node[title] (chal) at (\xl,-4.28) {Challenge solving};
\node[ask, below=0.6pt of chal] (chalq) {Can it be outsourced?};
\node[title] (sig) at (\xr,-4.28) {Verifier signal};
\node[ask, below=0.6pt of sig] (sigq) {What does the verifier observe?};

\draw[spine] (sigq.south) -- ++(0,-0.34);
\node[pill] (rq2) at (\xr,-5.48) {RQ2};
\node[ask, below=2.2pt of rq2] (rq2q) {What property matters?};
\draw[spine] (rq2q.south) -- ++(0,-0.34);
\node[title] (recon) at (\xr,-6.72) {Reconstruction};
\draw[spine] (recon.south) -- ++(0,-0.34);
\node[pill] (rq3) at (\xr,-7.55) {RQ3};
\node[ask, below=2.2pt of rq3] (rq3q) {What cost remains?};

\begin{pgfonlayer}{background}
  \fill[black!3.5, rounded corners=5pt]
    ([xshift=-14pt,yshift=6pt]inter.north west) rectangle
    ([xshift=8pt,yshift=-6pt]chalq.south east);
  \fill[black!3.5, rounded corners=5pt]
    ([xshift=-12pt,yshift=6pt]non.north west) rectangle
    ([xshift=14pt,yshift=-6pt]rq3q.south east);
\end{pgfonlayer}

\end{tikzpicture}%
}
\caption{An end-to-end bypass first splits by whether a challenge exists. RQ1 locates the bottleneck on each path (Section~\ref{sec:cloud_llm_results}). On the non-interactive path, RQ2 asks which property the verifier observes (Section~\ref{sec:recaptcha_behavior_vs_env}), and RQ3 asks what cost remains after that property is reconstructed (Section~\ref{sec:rq3_bypass}).}
\label{fig:argument_map}
\end{figure}

\noindent In this paper, we make the following contributions:

\noindent \textbf{First}, we locate the current bypass bottleneck by evaluating commercial solver services and LLM-based browser agents across interactive and non-interactive defenses. We find that commodity solver capabilities largely eliminate challenge-based defenses, while non-interactive defenses create a different bottleneck at the browser/session layer.

\noindent \textbf{Second}, we isolate this bottleneck using controlled browser configurations. We find that the defenses do not depend on a single generic notion of ``human-like behavior'': invisible reCaptcha~v2 and reCaptcha~v3 are strongly associated with a grounded Google account in authentic Chrome, whereas Turnstile is sensitive to the browser control path in our testbed. 

\noindent \textbf{Third}, we show the consequence of reconstructing the relevant trust property on our testbed. That grounded account transfers across agents and defense configurations, and once supplied, reCaptcha~v3 no longer determines whether the surrounding workflow succeeds or adds measurable marginal cost to an attempt.

Understanding whether this gap is closing matters beyond any single defense vendor. Large-scale scraping campaigns do not require solving a puzzle if the defense in front of them never raises one; they only require presenting an identity the defense is willing to trust. 
An end-to-end view reframes what ``defeating'' a non-interactive system means: not a single challenge broken once, but a trust signal that, once inherited or reconstructed, carries an agent through the workflow with no measurable defense-specific marginal cost per attempt. 
That absence of residual cost is what turns this into an economic problem, not just a technical one, as it lets cost-constrained adversaries mount far more attempts against far more targets. 
The calculus changes for defenders as well, as a generation of bot management built around raising the cost of solving a puzzle offers little protection once puzzle-solving is not where the decision is made.

%% file: 03-background.tex
\section{Background and Related Work}
\label{sec:background}


\subsection{Automated Traffic and Contemporary Defenses}

Automated web traffic spans a wide spectrum of intent. Beneficial bots underpin critical
infrastructure such as web crawlers, conversational agents, and monitoring
systems~\cite{khder2021web, kucherbaev2018human, yamanoue2017monitoring}, while malicious
automation drives abuse, including spam, credential stuffing, click fraud, and large-scale
vulnerability scanning~\cite{ferrara2019history, barkworth2022detecting, li2021good,
kondracki2022uninvited}. Li et al.~\cite{li2021good} found that 57\% of bots observed over seven months were plainly malicious, exploiting newly disclosed vulnerabilities within hours of public disclosure and evading detection via browser modification, spoofed User-Agent strings, and proxy routing.

Recent incident reports on account takeover and credential stuffing campaigns reveal the scale and structure of this abuse: Akamai recorded 193~billion credential stuffing attempts in 2024, a 66\% increase over the prior year~\cite{akamai2024soti}; analysis of 22 active credential-stuffing groups found that 65\% of attacks leveraged Captcha solver services and residential proxies despite 85\% of targeted organizations having bot detection in place~\cite{kasada2025ato}; and Verizon's 2025 DBIR corroborates this at the breach level, with credential stuffing constituting 19\% of daily authentication attempts in SSO logs and stolen credentials the initial access vector in 22\% of confirmed breaches~\cite{verizon2025dbir_stuffing}.

Understanding why these attacks persist requires examining the defenses they are designed to defeat.
Bot management systems remain a primary countermeasure for distinguishing human users from automated clients at web interaction points. 
These defenses can be categorized into two main groups. Challenge-based systems rely on explicit user interaction in image recognition or puzzle tasks and are often tuned by operators to balance user friction with abuse resistance. 
Non-interactive defenses, on the other hand, reduce explicit challenges and rely more on background telemetry. 
This shift changes the attack surface from solving puzzles to creating trusted signals for the verifier to trust the visitor. 

\subsection{Automated Solvers and Solving Services}

A substantial body of work demonstrates that the web bot management systems are vulnerable to automated solvers across all major challenge types. 
Text-based Captcha schemes have been broken using GANs, CNNs, and RNNs, with success rates exceeding 94\%~\cite{zhang2022counteracting, ye2018yet, tang2018research, dionysiou2020sok}. Image-based reCaptcha challenges have similarly been defeated, with deep learning-based solvers achieving 70\% accuracy by exploiting image repetition~\cite{sivakorn2016m}, and audio variants bypassed via off-the-shelf speech recognition~\cite{ieee11129490}. 
Even reCaptcha~v3's behavioral scoring has been compromised using reinforcement learning: Akrout~et~al.~\cite{akrout2019hacking} learn human-like mouse trajectories that earn a passing score (\(97.4\%\) under a grid-world formulation), and Tsingenopoulos~et~al.~\cite{tsingenopoulos2022captcha} use the v3 score itself as a reward signal, training over months of real interaction to reach evasion rates up to 99.6\%. Both attack the score through interaction behavior by learning mouse and timing dynamics. 
We later show that on our studied workflow, the score instead tracks persistent browser identity, which a hardened agent can supply without any behavioral training. 
Across all these approaches, the common bottleneck is data dependency, requiring each solver to be trained or fine-tuned as providers update their challenge pools.
Automated crawlers and solvers work together to bypass the defense; the crawler detects a Captcha challenge, delegates it to the solver, and resumes automated traversal upon receiving a solution or a valid token when required. In~\cite{ousat2024matter}, the authors create a pipeline to integrate a pre-trained text-based Captcha solver into a real-world crawler and achieve 63\% successful bypass in only five attempts for each target page.

In practice, most malicious actors are cost-driven and will gravitate toward whichever bypass method minimizes overhead and cost. Commodity solver services meet this demand by abstracting the problem entirely via hybrid human/automation pipelines~\cite{motoyama2010re, weng2019towards, gao2022demystifying, jin2023secure}, eliminating the need for any model expertise. Gao et al.~\cite{gao2022demystifying} estimate this underground market at \$4.8--128.1 million annually, and Jin et al.~\cite{jin2023secure} find that AI-powered solvers now rival human workers across major providers while services actively adapt to new challenge types.
The user simply submits a site key, page URL, and challenge type to an API, polls for a solution token, and injects it into the target form (sequence diagram available in Appendix~\ref{sec:appendix_solver_pipeline}). This capability is even integrated into open-source frameworks such as OpenBullet~2, making bypass a default feature of commodity credential-stuffing toolkits with no specialist expertise required~\cite{startupdefense2026stuffing}.


\subsection{LLM-Based Solvers and Browser Agents}

LLMs further reduce the cost and complexity of automated challenge solving. Deng~et~al.~\cite{deng2024oedipus} demonstrate an end-to-end framework that decomposes reasoning Captchas into LLM-solvable subproblems, achieving 63.5\% average success and adapting to previously unseen challenge designs. More recent work extends this capability to browser agents that autonomously perceive, reason about, and act on web content~\cite{yang2025agenticweb}. Teoh~et~al.~\cite{teohcaptchas} and Luo~et~al.~\cite{luo2025open} show that such agents can solve image-based challenges, including reCaptcha~v2 and hCaptcha. Wang~et~al.~\cite{wang2026cognition} further show that multimodal LLMs perform well on recognition-oriented visual challenges but struggle with fine-grained localization and counting; however, their offline evaluation measures perception against stored ground truth rather than access to a live verification service. Li~et~al.~\cite{li2025webcloak} similarly demonstrate the practical capability of LLM-powered agents to extract visual content from real-world webpages, reporting 88.7\% recall across 32 scraper variants.
These efforts are largely \emph{challenge-centric}: they improve an agent's ability to perceive or solve a puzzle once presented. Our focus is the upstream, end-to-end bypass problem: reaching the protected resource when the defense may instead reject the agent based on its browser environment, session identity, or control path, without presenting any challenge at all.

\begin{figure}[t]
\centering
\includegraphics[width=\linewidth]{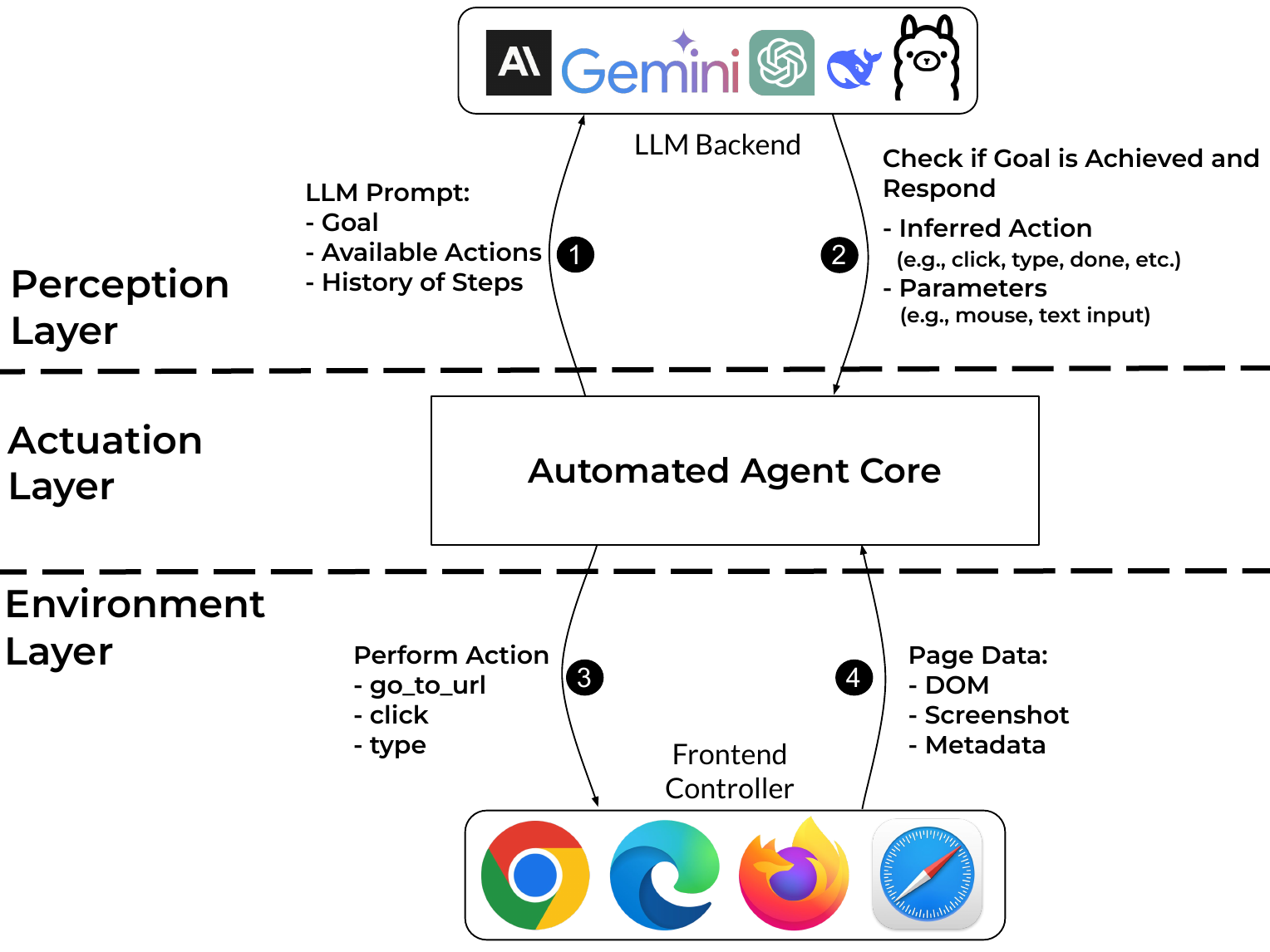}
\caption{LLM-based browser-agent workflow and three-layer architecture. The agent combines perception and reasoning by an LLM, browser actuation by the agent core, and a browser environment that provides observations to the agent and exposes browser and session properties to the verifier.}
\label{fig:llm_browser_agent_pipeline}
\label{fig:agent_stack}
\label{fig:agent_workflow_and_stack}
\end{figure}

Figure~\ref{fig:llm_browser_agent_pipeline} illustrates this workflow and decomposes an LLM-based browser agent into three layers. The perception and reasoning layer interprets browser observations and selects actions; the actuation layer executes those actions through the browser; and the environment layer provides the browser and session context in which the interaction occurs. A successful bypass must therefore satisfy both the capabilities required to complete the task and the properties observed by the verifier. Interactive defenses may engage all three layers by requiring the agent to perceive and solve a challenge, whereas non-interactive defenses can make their decision entirely from properties of the environment and interaction. We use this decomposition to identify where the effective bottleneck lies: challenge solving, environment acceptance, or the cost of reproducing the properties required for acceptance.

\section{Threat Model and Research Questions}
We define the threat model to clarify how the tools evaluated in this paper fit within an adversarial scenario and what capabilities are assumed on each side.

\subsection{Threat Model}
We model an economically motivated web attacker who seeks to bypass the bot management barrier to execute a downstream automated action, such as login attempts, form submissions, or large-scale
scraping.

\noindent\textbf{Attacker capabilities.} The attacker can: (i) call commercial solver APIs, (ii) run cloud or self-hosted LLM browser agents, (iii) modify open-source agent code, (iv) automate repeated trials under a fixed budget, (v) operate an authentic Chrome browser under a grounded consumer Google account, and (vi) route browser traffic through reputable residential or non-abusive IP addresses using commercial proxy or VPN.

\noindent\textbf{Attacker constraints.} The attacker has no privileged access to the target server, no Captcha secret keys, and no control over site-side verification logic, and operates exclusively through standard web interfaces and public APIs. 

\noindent\textbf{Defender model.} The defender deploys web bot management systems with standard backend verification and configurable thresholds.


\subsection{Research Questions}
\label{sec:rq}
This paper investigates how modern automation stacks can evade web bot management systems, including both interactive and non-interactive defenses. Prior work establishes that commercial solver services defeat challenge-based Captchas cheaply and reliably~\cite{weng2019towards,gao2022demystifying,jin2023secure}, but do not focus on the non-interactive, trust-scoring systems that now dominate deployment. 
We first locate where current automation fails (RQ1), then identify which session property creates that bottleneck (RQ2), and finally measure what cost remains once the property is reconstructed (RQ3).

\noindent \textbf{RQ1. Where is the bottleneck?}
Across commercial solver services and LLM-based browser agents, which stage of the end-to-end automation workflow limits bypass against modern bot-management systems?

\noindent \textbf{RQ2. What creates the bottleneck?}
For non-interactive defenses, which session property---browser authenticity, account identity, or the agent control path---dominates the verification outcome, and does supplying that property reverse the verdict?

\noindent \textbf{RQ3. What remains after reconstruction?}
Once the relevant trust property has been reconstructed, does the defense impose any residual capability or per-attempt cost on an adversary?

%% file: 04-methodology.tex

\section{Experimental Testbed}
This section details the testbed environment and the systems under evaluation, including the selected defenses, commercial services, and LLM-based agents.

\subsection{Controlled Web Bot-Mitigation Testbed}

We constructed a controlled public testbed consisting of seven protected web applications, each implementing a standard HTML form protected by one of the defense configurations listed below and deployed on a dedicated subdomain and configured with a distinct bot-mitigation setting.
This controlled setup allows us to monitor interactions in isolation without affecting public services.

\label{sec:selected_captcha}
\noindent \textbf{hCaptcha}~\cite{hcaptcha_about} is deployed in two separate difficulty tiers sharing a common interface: an \textit{Easy} tier of unambiguous object-recognition tasks, and a \textit{Hard} tier of fine-grained image classification and puzzles that raise resistance to automated solvers while preserving human solvability, at higher outsourcing cost.

\noindent \textbf{reCaptcha v2}~\cite{recaptcha_v2_about} shares a risk-scoring engine across two variants. The \textit{Checkbox} variant requires an explicit ``I'm not a robot'' click and surfaces an image-grid challenge when background signals are inconclusive; the \textit{Invisible} variant shows no widget, verifying in the background and triggering a grid challenge only when the risk score is high.

\noindent \textbf{reCaptcha v3}~\cite{recaptcha_v3_about} operates without any visible challenge or user interaction, silently assigning a continuous risk score between 0.0 and 1.0 to each session from behavioral and environmental signals, with no puzzles to solve. Sites use this score to gate access or trigger step-up authentication, with a default acceptance threshold of 0.5~\cite{google_recaptcha_v3}. 

\noindent \textbf{Cloudflare Turnstile}~\cite{cloudflare_turnstile} is deployed in \textit{Managed} (checkbox) and \textit{Invisible} modes, which differ mainly in whether the widget is shown. Both run background integrity checks on the browser environment rather than explicit challenge solving, placing Turnstile closer to reCaptcha~v3 as a non-interactive defense.

All experimental pages have been publicly accessible for over eight months before the final evaluations and received regular benign interactions multiple times per week. Maintaining long-lived deployments exposed the systems to realistic background traffic, including potential automated scanning and unsolicited access attempts commonly observed on publicly reachable web services. This extended deployment period also reduced the likelihood of evaluating freshly initialized or insufficiently calibrated configurations, which is particularly important for behavioral and risk-based systems such as reCaptcha v3 that rely on historical telemetry and reputation signals for risk assessment.


These bot management systems were selected to span the full spectrum of defense paradigms, from explicit challenge-response mechanisms (reCaptcha v2, hCaptcha) to non-interactive verification (Turnstile, reCaptcha v3), enabling a comprehensive assessment of each attacker stack's capabilities and limitations. Screenshots of the deployed login pages are shown in Figure~\ref{fig:sample_login_v2_hCaptcha} of Appendix~\ref{sec:appendix_testbed_pages}.
To verify that the set of defenses we evaluate reflects real-world deployment patterns, we conducted a preliminary prevalence analysis using the Chrome User Experience Report (CrUX) dataset~\cite{google_crux}, which lists the landing-page URLs of the web's most-visited sites. We were able to successfully crawl 350{,}174 of these landing pages and inspected each page's script tags for signatures of the four major bot-management services. Because Captcha widgets are typically deployed on specific interior pages, such as sign-up, sign-in, or checkout forms, rather than on the landing page itself, this crawl represents a lower bound on true deployment prevalence. Even so, we identified 35{,}786 sites embedding at least one of these services. We observed that reCaptcha v3 accounts for 24{,}073 sites, over two-thirds (67\%) of all detected deployments, ahead of reCaptcha v2 (8{,}908; 25\%), Cloudflare Turnstile (1{,}642; 5\%), and hCaptcha (1{,}164; 3\%). Details of this experiments is provided in Appendix \ref{sec:appendix_prevalence}.
This distribution confirms that non-interactive, score-based defenses, and reCaptcha v3 in particular, are not a niche configuration but the dominant deployment choice in practice, which motivates our focus on not only challenge-based defenses, but also the non-interactive ones.

It is worth noting that since we conducted our evaluation, Google has consolidated reCaptcha v2 and v3 under a new umbrella product, Google Cloud Fraud Defense~\cite{google_fraud_defense}, which folds the classic checkbox/invisible-challenge and score-based APIs into a broader fraud, account-protection, and transaction-defense platform integrated with Google Cloud. The verification flow we analyze in this paper is preserved under the new branding, meaning that the threat model and attack surface remain applicable to the migrated service. We frame our results as characterizing the reCaptcha v2/v3 defenses as broadly deployed at evaluation time, while noting the Google's rebranding to Fraud Defense.

\subsection{Selected Solving Services}
We selected a representative set of services based on their prevalence and ecosystem footprint. Selection was guided by public web-traffic signals and integration breadth, prioritizing services with high user activity commonly deployed in automation pipelines~\cite{motoyama2010re,weng2019towards,gao2022demystifying,jin2023secure}. As summarized in Table~\ref{tab:captcha-services}, the chosen services span a monthly user base of 190K–1.2M and collectively support a variety of Captcha types via human labor, machine-learning models, or hybrid approaches~\cite{motoyama2010re,jin2023secure}. Pricing ranges from \$0.10 to \$5.00 per 1,000 solves and tracks the operational complexity of the supported defense.

\begin{table}[t]
\centering
\small
\caption{Commercial solving services and their popularity. User volume and ranking extracted from SimilarWeb~\cite{similair_web}.}
\label{tab:captcha-services}
\resizebox{\columnwidth}{!}{%
\begin{tabular}{llll}
\toprule
\textbf{Service} & \textbf{Monthly Users} & \textbf{Global Rank} & \textbf{Price/1K} \\
\midrule
2Captcha              & 1.2M  & 37,810  & \$1.45--\$2.99 \\
Anti-Captcha          & 450K  & 84,392  & \$2.00--\$5.00 \\
AZCaptcha             & 190K  & 145,771 & \$0.10--\$1.00 \\
BestCaptchaSolver     & 300K  & 110,531 & \$1.80--\$3.00 \\
Capsolver             & 650K  & 72,405  & \$1.00--\$3.00 \\
CapMonster            & 980K  & 49,219  & \$0.60--\$1.30 \\
ImageTyperz           & 210K  & 132,884 & \$1.80--\$2.10 \\
\bottomrule
\end{tabular}
}%
\end{table}


\subsection{Selected LLM-Based Browser Agents}
\label{sec:selected_agents}

To assess the real-world capabilities of modern large language model (LLM)-driven browser agents in bypassing bot management solutions, we selected a diverse set of representative systems spanning both research prototypes and production deployments. Our selection prioritizes agents capable of multi-step web reasoning and interaction, while ensuring reproducibility through publicly accessible implementations. We further emphasize architectural diversity across DOM-based, vision-based, and hybrid designs, as well as variation in execution environments ranging from cloud-hosted services to fully self-hosted and browser-native systems.

We organize the evaluated systems into four deployment settings: cloud-hosted agents, self-hosted agents, AI-assisted browsers, and browser-extension-based agents. 
RQ1 focuses on the cloud-hosted, AI-assisted-browser, and extension deployments as the off-the-shelf attacker surface. Self-hosted deployments are reserved for RQ2 and RQ3, where they serve as the instrumented vehicles for isolating and reconstructing account identity as they expose execution traces and can be bound to custom browsers.

\noindent \textbf{Cloud Agents.} We evaluate cloud-hosted deployments of \textsc{Browser-Use}~\cite{browseruse2024}, \textsc{Skyvern}~\cite{skyvern2024}, and \textsc{Manus}~\cite{meta_manus}. \textsc{Browser-Use} integrates structured DOM interaction with visual grounding; \textsc{Skyvern} combines DOM extraction and visual context, and its commercial deployment includes a dedicated Captcha-solving pipeline~\cite{skyvern_captcha_blog}; \textsc{Manus} is a commercial end-to-end autonomous browsing system. Collectively, these reflect a low-barrier threat model in which adversaries rely solely on hosted agent interfaces.

\noindent \textbf{Self-Hosted Agents.} We evaluate self-hosted \textsc{Browser-Use}, \textsc{Skyvern}, \textsc{OpenManus}~\cite{openmanus2025}, and \textsc{SeeAct}~\cite{zheng2024seeact}, which expose internal prompts and execution traces for fine-grained analysis and include no proprietary integrations or external solving services. 

\noindent \textbf{AI-Assisted Browsers.} We evaluate \textsc{BrowserOS}~\cite{browseros2025} and \textsc{Perplexity Comet}~\cite{perplexity2025comet}, which embed LLM capabilities directly into the browsing interface rather than driving a separate automation stack: \textsc{BrowserOS} integrates reasoning with native browser operations, and \textsc{Comet} is a browser-integrated conversational assistant.

\noindent \textbf{Browser-Extension Agents.} Lastly, we evaluate \textsc{NanoBrowser}~\cite{nanobrowser2024}, a browser-extension agent operating directly within a standard user browser environment. Unlike instrumented automation frameworks, it inherits native browser state---cookies, history, extensions---substantially reducing fingerprinting artifacts, which helps isolate whether failures arise from reasoning limitations or environmental detectability.


\subsection{Live Evaluation Requirements}

Restricting the study to offline-reproducible challenges would exclude precisely the defenses that dominate in practice. This shapes our methodology directly: throughout the paper, every outcome we report is obtained against a live deployment rather than a static challenge corpus, and for a substantial part of our analysis this is a requirement rather than a preference. Non-interactive defenses never present a challenge and therefore produce no artifact that can be collected, annotated, and replayed offline; their verdict is a score computed over environmental and behavioral signals observed during the session itself and returned as a token, so the measurement exists only when a live provider evaluates a live session. Invisible reCaptcha v2 runs the same kind of silent check first, but, unlike v3 and Turnstile, can escalate into an interactive grid challenge when that check is inconclusive. Because whether it escalates is itself decided by live environmental and behavioral signals, that decision is no more offline-reproducible than the score Turnstile or v3 returns, even though the resulting grid, once rendered, looks like an ordinary challenge.

Live evaluation does constrain scale. Each attempt consumes a real challenge from a production service, is subject to provider-side rate limiting and adaptive scoring, incurs agent and model cost, and must be verified against the resulting page state rather than a stored label, so our trial counts are modest relative to dataset-based solver benchmarks that evaluate hundreds of instances per task type~\cite{wang2026cognition, luo2025open}. We regard this as a favorable trade rather than a shortfall. Even for interactive challenges, where an offline formulation does exist, a dataset-based benchmark holds constant the authenticity, behavioral, and token-validation layers that convert a solved challenge into access, and so measures only what a solver could achieve in isolation, bounded away from the deployment context that ultimately decides acceptance. Each outcome we report is instead an actual verification decision: a token issued for a browser the provider observed while it solved, then accepted or rejected by the vendor's own validation endpoint. This distinction is what Sections 5–7 follow up on.

%% file: 05-experiments.tex
\section{RQ1: Off-the-Shelf Effectiveness}
\label{sec:cloud_llm_results}

RQ1 asks where modern bot defenses stop today's off-the-shelf automation stacks. We evaluate the two attacker classes, commercial solver services and LLM-based browser agents, against the same deployed defenses. 
The purpose is not to re-establish that challenge-based Captchas can be solved, which prior work has already shown, but to identify the point in the end-to-end workflow for both interactive and non-interactive defenses.

\subsection{Solver Services Against Modern Defenses}
\label{sec:solver-service-eval-pipeline}







\noindent The effectiveness of solver services against challenge-based Captchas is previously studied~\cite{motoyama2010re,weng2019towards,gao2022demystifying,jin2023secure}. We specifically measure this under our current deployments to confirm the effectiveness on challenge-based defenses and to further extend the measurement to the non-interactive systems that have not been a main focus of the prior work.

The measurement is a per-service worker loop that visits the page, requests a verifiable token from the solvers, injects the token into a form, and attempts to log in (Figure~\ref{fig:solver_measurement}). We assign each of the seven services a dedicated worker. For every testbed page the service claims to support, the worker builds a task from the page URL, site key, challenge type, and a minimum-score parameter for reCaptcha v3, submits it through the service API, and polls until a token arrives or a 4-minute timeout expires. A returned token is not counted on the solver's say-so: we send it to the vendor's own verification endpoint (hCaptcha \texttt{siteverify}, Google's reCaptcha API, Cloudflare Turnstile's endpoint) for final verdict. 
Challenge-based tokens are accepted or rejected; a v3 token succeeds only if the returned score exceeds Google's default threshold of 0.5~\cite{google_recaptcha_v3}. We record the verified outcome and the wall-clock latency and repeat until 100 completed trials per service--target pair. 
Per-service results are provided in ~\ref{tab:solver-performance-full} of Appendix~\ref{sec:appendix_solver_pipeline}

\begin{figure}[t]
\centering
\resizebox{\columnwidth}{!}{%
\begin{tikzpicture}[
  font=\small,
  card/.style={
    draw, line width=0.85pt, rounded corners=2.5pt, align=center,
    inner sep=3.5pt, minimum width=2.05cm, minimum height=1.45cm
  },
  arr/.style={-{Latex[length=1.7mm]}, line width=0.9pt},
  chip/.style={
    draw, rounded corners=1.8pt, fill=white, inner sep=2pt,
    font=\scriptsize, align=center, line width=0.5pt
  },
  num/.style={
    circle, fill=black, text=white, font=\scriptsize\bfseries,
    inner sep=0.9pt, minimum size=1em
  },
]
\node[card] (worker) at (0, 2.15)
  {\large\faCogs\\[1.5pt]
   \textbf{Worker}\\[-1pt]
   {\scriptsize\faCheck\ verdict\ \ \faClockO\ time}};
\node[card] (solver) at (-2.15, -1.25)
  {\large\faCloud\\[1.5pt]
   \textbf{Solver}\\[-1pt]
   {\textbf{Service}}};
\node[card] (web) at (2.15, -1.25)
  {\large\faGlobe\\[1.5pt]
   \textbf{Website}\\[-1pt]
   {\scriptsize \texttt{siteverify}}};

\draw[arr] (worker.west) to[out=180, in=90, looseness=0.5] (solver.north);
\draw[arr] (solver.east) to[out=40, in=245, looseness=0.1] ($(worker.south)-(0.3,0)$);
\draw[arr] (worker.east) to[out=0, in=90, looseness=0.5] (web.north);
\draw[arr] (web.west) to[out=140, in=-65, looseness=0.1] ($(worker.south)+(0.3,0)$);

\node[chip] (ch1) at (-3, 0.85)
  {\faKey\ \texttt{site\_key}, \texttt{URL}, \texttt{type}};
\node[num, left=2pt of ch1] {1};
\node[chip] (ch2) at (-0.6, -0.15)
  {\faTicket\ \texttt{token}};
\node[num, left=2pt of ch2] {2};
\node[chip] (ch3) at (2.3, 0.85)
  {\faTicket\ \texttt{token}};
\node[num, right=2pt of ch3] {3};
\node[chip] (ch4) at (0.7, -0.15)
  {\faCheck\ \texttt{verdict}};
\node[num, right=2pt of ch4] {4};

\end{tikzpicture}%
}
\caption{One trial of the solver-service measurement. The worker submits the challenge parameters to the commercial solver~(\protect\circled{1}), receives a token~(\protect\circled{2}), submits the token to the protected website~(\protect\circled{3}), and records the resulting verdict and wall-clock latency~(\protect\circled{4}).
}
\label{fig:solver_measurement}
\end{figure}

Table~\ref{tab:solver-performance} summarizes the results for this experiment. When a solver service supports the Captcha type, every challenge-based defense falls completely: all seven providers return a vendor-accepted token on all 100 reCaptcha~v2 trials, and the six that offer Turnstile do the same. The solve times, however, span more than an order of magnitude, and that spread is the first visible difference in how these services operate. CapMonster and Capsolver return v2 and Turnstile tokens in 4--11\,s; 2Captcha is in the same band on those types (7--14\,s). Anti-Captcha, ImageTyperz, and BestCaptchaSolver take 40--90\,s for the same 100\% outcome, and AZCaptcha takes 102--132\,s on v2. Sub-10-second solves are the signature of an automated token pipeline. The solve times in the tens of seconds are consistent with a human farm or a real browser sitting behind the API. The client interface is identical in every case (submit, poll, inject), so an adversary does not have to know which backend they are using.

\begin{table}[t!]
\centering
\small
\setlength{\tabcolsep}{4pt}
\renewcommand{\arraystretch}{1.15}
\caption{Third-party solver performance, summarized over the seven evaluated services. \textbf{Services}: number of services supporting the type; \textbf{Succ.}\: range of success rates across supporting services; \textbf{Time}: median solve time. Challenge-based defenses are solved nearly perfectly, while success against reCaptcha v3 ranges from 0--63\% (23\% on average). 
}
\label{tab:solver-performance}
\begin{tabular}{llccc}
\toprule
\textbf{Captcha} & \textbf{Variant} & \textbf{Services} & \textbf{Succ.\ (\%)} & \textbf{Time (s)} \\
\midrule
\multirow{2}{*}{\textbf{hCaptcha}}
 & Easy & 2/7 & \textbf{100} & 54.7 \\
 & Hard & 2/7 & 58--100 & 92.6 \\
\midrule
\multirow{2}{*}{\textbf{reCaptcha v2}}
 & Checkbox & 7/7 & \textbf{100} & 49.4 \\
 & Invisible & 7/7 & \textbf{100} & 52.7 \\
\midrule
\textbf{reCaptcha v3} & Score ($>$0.5) & 7/7 & \textit{0--63} & 17.5 \\
\midrule
\multirow{2}{*}{\makecell{\textbf{Cloudflare}\\\textbf{Turnstile}}}
 & Managed & 6/7 & \textbf{100} & 13.8 \\
 & Invisible & 6/7 & \textbf{100} & 17.0 \\
\bottomrule
\end{tabular}
\end{table}

hCaptcha is the exception in coverage. Only 2Captcha and ImageTyperz offer hCaptcha solving. Easy-mode tokens verify at 100\% for both (median 54.7\,s); on the hard tier, ImageTyperz stays at 100\% (82.7\,s) while 2Captcha drops to 58\% (102.5\,s). That gap, and the \$1.80--\$2.99 per 1{,}000 price of the two providers that support it (Table~\ref{tab:captcha-services}), is the cost of a challenge that still needs human annotation or a fine-grained vision pipeline rather than a token minted from a site key. 
AZCaptcha, the cheapest service at \$0.10--\$1.00, supports neither hCaptcha nor Turnstile. Across the types a service does support, prices sit between \$0.10 and \$5.00 per 1{,}000 solves.


The other exception in the data is reCaptcha~v3 verdicts. All seven services accept v3 tasks, but three (AZCaptcha, CapMonster, and Capsolver) produce no token that clears 0.5, and they do so quickly (1.6--17.5\,s): the API returns the token; the score is just not high enough. 
The other four clear the threshold only intermittently (2Captcha 26\%, Anti-Captcha 30\%, ImageTyperz 41\%), and only BestCaptchaSolver reaches a majority, at 63\%, taking 39.7\,s per attempt. 
Those slower v3 times sit in the same band as the slower challenge-based solvers, which is consistent with a provider that presents a real browser rather than minting a token inside its own infrastructure. 
A token minted off-session carries little of the history the score is computed over, so outsourcing cannot reach a defense that scores the session itself.

Solver outsourcing is highly effective when the defense's decision is attached to a solvable artifact, but becomes substantially less effective when the verifier scores the session that produced the token. The solver results therefore reveal a boundary, but not its cause: against reCaptcha~v3, the returned token alone does not appear to carry the trusted session context required by the verifier. The only provider that partially succeeds is consistent with a solver that presents a real browser profile behind a residential exit. This motivates a different question for autonomous browser agents: unlike an API solver, an agent controls the browser that generates the session and can therefore potentially supply the missing session context directly.

\begin{figure}[t]
\centering
\includegraphics[width=\columnwidth]{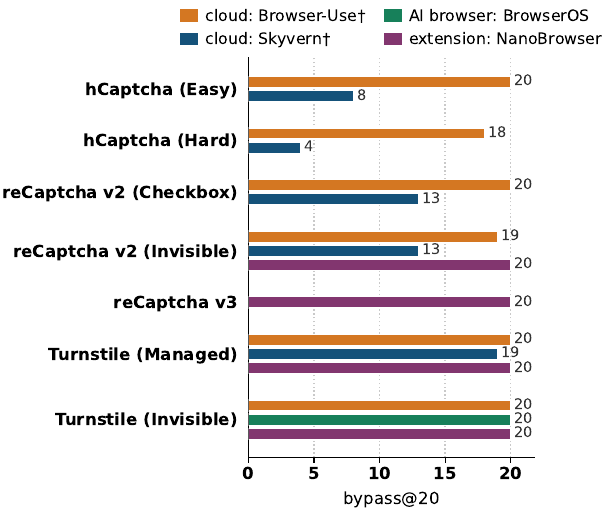}

\caption{
Off-the-shelf agent bypass performance across defenses, measured as \textit{bypass@20}. Missing bars denote $0/20$ successful bypasses. Manus achieved $0/20$ on every defense and is omitted for clarity. Comet declined all trials under its safety policy and is omitted. $^\dagger$Browser-Use~\cite{browseruse_bot_detection} and Skyvern~\cite{skyvern_captcha_blog} document proprietary Captcha-solving pipelines.
}
\label{fig:combined_agent_eval}
\end{figure}

\subsection{Agents Against Modern Defenses}
\label{sec:rq1_protocol}

We now turn to the off-the-shelf LLM-based browser agents, including cloud-hosted, AI-browser, and extension deployments introduced in Section~\ref{sec:selected_agents} and evaluate them under their default configurations. 
For this experiment, each agent is provided with a simple task prompt (Figure~\ref{fig:cracking_prompt} in Appendix~\ref{sec:appendix_prompt}) and is asked to log in to each of the deployed websites by solving the challenge if faced any.
A successful attempt requires the agent to both satisfy the Captcha verification mechanism (solving an interactive challenge or achieving acceptance in a non-interactive system) and submit the form; failing either is a failure. We manually verify every outcome against the resulting final page rather than the agent's self-report.

We evaluate each configuration using a \emph{bypass@20} metric, defined as the number of successful bypasses over 20 independent attempts. This metric captures both whether a configuration can bypass the defense and how reliably it does so across repeated trials.


The results, provided in Figure~\ref{fig:combined_agent_eval}, reveal that successful bypass is not determined solely by the reasoning capabilities of the underlying LLM. 
Different defense systems pressure distinct capabilities, including challenge-solving, execution-environment authenticity, and behavioral analysis.
Notably, Comet declined to fill in or submit login forms, citing concerns about bypassing authentication flows. This suggests that Comet's failures reflect an intentional safety policy rather than a capability gap, and its bypass results should be interpreted accordingly.

\subsection{Agents vs.\ Interactive Defenses}
\label{sec:rq1_interactive}

Interactive systems, including reCaptcha v2 and hCaptcha, primarily pressure the perception and solver capabilities of autonomous agents. Cloud-based agents can frequently solve the challenge when faced with image grids and simple to medium puzzles.
Some agents, including Manus and BrowserOS, frequently identify the verification mechanism and interact with it, yet fail to complete the challenge, indicating they can locate the Captcha interface but lack the specialized perception and solver capabilities to satisfy it.
Bypass here depends less on planning ability and more on access to specialized challenge-solving integrations. 
This is clearest in the cloud-hosted Browser-Use and Skyvern deployments that are able to bypass hCaptcha and reCaptcha v2 variants, consistent with their documented in-house Captcha-solving pipelines~\cite{browseruse_bot_detection,skyvern_captcha_blog}. Agents without that integration identify the widget yet fail the puzzle. 

Figure~\ref{fig:hcaptcha_samples} in the appendix shows representative hCaptcha puzzles that agents were not able to clear. 
Recent offline evaluations help explain why hard hCaptcha variants remain unsolved. Wang et al.~\cite{wang2026cognition} benchmark seven multimodal LLMs on static Captcha images and find that coarse recognition is largely solved, while six families that require precise localization, ordered multi-object binding, or counting under clutter stay below 20\% Pass@1 under direct prompts, optimized prompts, and few-shot demonstrations. 
Failures stem from spatial grounding: models often describe the correct target yet emit coordinates outside the verifier's tolerance. 
The puzzles in Figure~\ref{fig:hcaptcha_samples} depend on exactly these skills, so the agent failures we observe are consistent with a model-level limit rather than with an artifact of our live setup.
Additionally, frameworks such as~\cite{deng2024oedipus} are already lowering the barrier to build solvers to bypass such challenges, which can potentially be integrated into browser agents.

\subsection{Agents vs.\ Non-Interactive Defenses}
\label{sec:rq1_noninteractive}

Non-interactive defenses, such as Cloudflare Turnstile and reCaptcha v3 present no explicit puzzle, so challenge solving is no longer the immediate verification bottleneck; instead, the verifier makes its decision from properties observable from the agent.
Turnstile does not uniformly reject automation; certain configurations, particularly cloud-hosted Browser-Use and NanoBrowser, bypass it successfully as we observe the Turnstile checkbox being automatically marked and the invisible check verified upon request. For Skyvern, we noticed that while the Turnstile checkbox is not automatically completed, the agent can consistently interact with it and mark it manually in almost every run, resulting in a bypass for this mode.

Invisible reCaptcha v2 is a hybrid rather than a third member of this category. It runs the same kind of silent trust check first (Section~\ref{sec:selected_captcha}), and an environment trusted enough clears it the same way Turnstile and v3 are cleared: NanoBrowser bypasses it in every trial (Figure~\ref{fig:combined_agent_eval}) without triggering the grid.
When the silent check is inconclusive, however, invisible v2 falls back to the ordinary image-grid challenge, and we score that fallback exactly as we score the Checkbox variant: triggering the grid is not itself a failure, only failing to solve it is. Cloud-hosted agents then clear that fallback through their dedicated solver pipeline.
reCaptcha v3 shows the strongest dependence on execution-environment authenticity, silently scoring browser, environmental, and behavioral signals rather than posing a visible challenge. The browser-extension deployment (NanoBrowser) is the only configuration that consistently clears it: it operates inside a real user Chrome signed into a consumer account. This observation motivates the analysis of  Section~\ref{sec:recaptcha_behavior_vs_env}, where we investigate the agents and browser attributes in more detail.



\begin{tcolorbox}[
colback=gray!10,
colframe=black!40,
title=\textbf{Key Observation},
fonttitle=\bfseries,
]
Across both off-the-shelf attacker classes, successful bypass is determined by the availability of specialized capabilities rather than by agent reasoning, and the two classes fail with similar patterns. Solver services clear challenge-based Captchas at near-perfect rates at low per-solve cost but collapse against reCaptcha v3, likely because a token minted in a provider's infrastructure carries little trust signals. Challenge-based systems, including invisible reCaptcha v2 once it escalates into a grid, primarily depend on solver integrations, while the purely non-interactive systems (Turnstile, reCaptcha v3) rely on trust signals the agent carries, not reasoning or task execution. 
\end{tcolorbox}

The RQ1 results identify a change in the attack surface rather than simply a difference in success rate: when no explicit challenge is presented, the bottleneck moves from challenge solving toward properties of the agent and the browser session. RQ2 provides a more detailed analysis of those properties which determine the verifier's decision.

\section{RQ2: Isolating the Environment Signals}
\label{sec:reCaptcha_v3}
\label{sec:recaptcha_behavior_vs_env}

We established that the difficult part of non-interactive bypass is not completing the workflow itself, but obtaining a session that the verifier accepts. Here we ask what property of that session creates the bottleneck. A non-interactive verifier can score the browser and its session at three nested levels: whether the execution environment exhibits automation artifacts, whether the session is bound to a grounded, persistent consumer identity rather than a bare browser, and the control path through which an agent's high-level actions are translated into browser-visible input and DOM operations. We isolate the first two using Chrome incognito, a cold profile, and a full profile account, which separate a merely authentic browser from one carrying real account history and engagement.


We hold the agent fixed, and with it its browser control path, and walk the browser through a ladder that turns authenticity on (instrumented Chromium, then Chrome incognito or a cold profile) and then account identity on (a full profile).

\subsection{Agents Across Browser Configurations}
\label{sec:rq2_env_grid}
\label{sec:rq2_env_sublayers}
\label{sec:rq2_reconstruction}

Analyzing the details of the interactions requires a stack whose browser we can replace. A cloud agent ships a fixed browser, so the agent and environment stay fused. We therefore use the open-source, self-hosted agents reserved for this purpose in Section~\ref{sec:selected_agents} (i.e., Browser-Use, Skyvern, OpenManus, and SeeAct). These stacks are able to attach to the Chrome DevTools Protocol (CDP)~\cite{chrome_devtools_protocol}, expose their execution path without a vendor-side Captcha pipeline, and let us hold the agent fixed while varying only the browser.
The browser configurations we analyze for this experiment include:

\noindent \textbf{Instrumented.} The agent's default stack: bundled Chromium (e.g., Playwright), a fresh context, and the automation artifacts that the stack exposes.

\noindent \textbf{Chrome Incognito.} A genuine Google~Chrome binary in an incognito window, driven by the same agent over CDP. This turns authenticity on while keeping account identity off, modulo incognito's own cookie-policy and storage-partitioning behavior.

\noindent \textbf{Chrome Cold Profile.}
The same Google~Chrome binary launched with a newly initialized, throwaway browser profile and signed in to a newly created consumer Gmail account. 
The newly created account provides an authenticated identity, but has minimal, artificially seeded history and is not associated with the long-lived browser state used in the full profile condition. This configuration therefore isolates an authenticated but cold browser account state.

\noindent \textbf{Chrome Full Profile.}
The same Google~Chrome binary launched with a persistent browser profile associated with an established consumer Gmail account. Unlike the cold profile, the account and its browser environment have an existing history of ordinary use and persistent first-party Google-service state. The profile contains accumulated cookies, site storage, browser history, and other state resulting from prior use. 
We use the established account and its associated persistent state as the defining properties of this configuration, rather than attributing the observed behavior to any individual cookie, storage object, or browser artifact.


\begin{table}[t]
\centering
\footnotesize
\renewcommand{\arraystretch}{1.15}
\setlength{\tabcolsep}{3.5pt}
\definecolor{envno}{HTML}{C46A63}
\definecolor{envok}{HTML}{1B7F3B}
\DeclareRobustCommand{\envY}{\textcolor{envok}{\ding{51}}}
\DeclareRobustCommand{\envN}{\textcolor{envno}{\ding{55}}}
\caption{Verification outcomes across agent--defense--browser configurations.
Each cell summarizes 20 trials: \texorpdfstring{\envY}{Y} denotes successful admission in 18--20 trials,
where the remaining trials, when present, failed because of agent workflow or execution
errors rather than verifier rejection; \texorpdfstring{\envN}{N} denotes 0 admissions in 20 trials.
The shaded heading names the agent's input path.}
\label{tab:rq2_env_grid}
\begin{tabular}{lcccc}
\toprule
& \textbf{Instrumented}
& \makecell{\textbf{Chrome}\\\textbf{incognito}}
& \makecell{\textbf{Chrome}\\\textbf{cold profile}}
& \makecell{\textbf{Chrome}\\\textbf{full profile}} \\
\midrule
\rowcolor{black!6}
\multicolumn{5}{l}{\textbf{Browser-Use}\enspace{\normalfont\scriptsize\textcolor{black!45}{Synthesizes keyboard and mouse events}}} \\
Invis. reCaptcha v2 & \envN & \envN & \envN & \envY \\
reCaptcha v3        & \envN & \envN & \envN & \envY \\
Invis. Turnstile    & \envY & \envY & \envY & \envY \\
\addlinespace
\rowcolor{black!6}
\multicolumn{5}{l}{\textbf{Skyvern}\enspace{\normalfont\scriptsize\textcolor{black!45}{Framework keystrokes and mouse clicks}}} \\
Invis. reCaptcha v2 & \envN & \envN & \envN & \envY \\
reCaptcha v3        & \envN & \envN & \envN & \envY \\
Invis. Turnstile    & \envN & \envY & \envY & \envY \\
\addlinespace
\rowcolor{black!6}
\multicolumn{5}{l}{\textbf{OpenManus}\enspace{\normalfont\scriptsize\textcolor{black!45}{Scripted click with no mouse path}}} \\
Invis. reCaptcha v2 & \envN & \envN & \envN & \envN \\
reCaptcha v3        & \envN & \envN & \envN & \envY \\
Invis. Turnstile    & \envN & \envN & \envN & \envN \\
\addlinespace
\rowcolor{black!6}
\multicolumn{5}{l}{\textbf{SeeAct}\enspace{\normalfont\scriptsize\textcolor{black!45}{Screenshot-selected target, then a framework mouse click}}} \\
Invis. reCaptcha v2 & \envN & \envN & \envN & \envN \\
reCaptcha v3        & \envN & \envN & \envN & \envY \\
Invis. Turnstile    & \envN & \envY & \envY & \envY \\
\bottomrule
\end{tabular}
\end{table}

Table~\ref{tab:rq2_env_grid} is the factorial design the ladder is built to produce.
Each row holds the agent, and therefore its control path, fixed.
Each successive column then turns on browser authenticity (instrumented to incognito or cold) and, last, persistent identity (full profile).
A verdict that changes along a row is therefore an environment effect: the agent, the task, and the input transport are unchanged.
A verdict that differs across rows in the same column is a control-path effect: the browser is the same, and the agents are not. We notice that the defenses do not score the same property.

Invisible reCaptcha~v2 admits Browser-Use and Skyvern only in the last experiment with authentic Chrome bound to an established account.
The two preceding authentic-Chrome cells, incognito and cold, remain rejected, so a genuine binary without that account state is not enough.
OpenManus and SeeAct remain rejected even with the full profile, so that account state is also not sufficient for every control path.
reCaptcha~v3 occupies the same last-column pattern, and we return to its scores below.
reCaptcha~v3 shows the strongest dependence on the established profile.
All non-full-profile configurations are rejected and sit at a score of $0.1$; full-profile configurations achieve $18$--$20/20$ admissions with scores from $0.7$ to $1.0$ on successful runs.
The few failed full-profile v3 runs are not verifier rejections: the agent does not finish the login workflow.
Once the established profile is supplied, residual differences across agents are workflow reliability, not a different v3 verdict.

Turnstile flips one step earlier, if it admits at all.
Skyvern and SeeAct are refused only while they drive their bundled Chromium, and are admitted on every genuine Chrome configuration, including incognito and cold profiles that carry no Google account. Browser-Use is already admitted in the instrumented run, and OpenManus is admitted in none.
Turnstile therefore tracks authenticity for two agents, is insensitive to both authenticity and identity for Browser-Use, and remains closed for OpenManus regardless of the environment.
The established profile can change a verdict, but it is not the quantity every defense is measuring.

These contrasting outcomes under the same browser environment motivate a closer look at how each agent turns a planned action into browser-visible input.
The control path is a property of the agent, not of the Chrome configuration: we hold it fixed while walking the authenticity and identity ladder, so a change in verdict along that ladder cannot be attributed to a change in how the agent types or clicks.
We group the four agents into three control-path categories, from the most complete synthesis of user input to the least.

\noindent\textbf{Protocol-level event synthesis.}
Browser-Use drives the page directly at the browser's automation protocol, dispatching individual key and pointer events so that the page observes a full event sequence for every keystroke and click, regardless of which Chrome binary it is attached to.
This is the only agent admitted by Invisible Turnstile in the instrumented configuration: its input transport already resembles an authentic session closely enough that swapping in genuine Chrome changes nothing.

\noindent\textbf{Framework-level interaction primitives.}
Skyvern and SeeAct's click actuation act through a browser-automation framework's locator API rather than the raw protocol.
This still produces a plausible keystroke or pointer event sequence, but the framework's own automation context leaves artifacts when it is running the bundled Chromium binary that ships with the agent.
Both are refused by Turnstile in the instrumented configuration and admitted as soon as the same control path is pointed at a genuine Chrome install, including an incognito window with no account history.
For these two agents, authenticity of the browser binary, not persistent identity, is what flips the verdict.

\noindent\textbf{Programmatic value assignment.}
OpenManus, and SeeAct's own text-entry path, write field values directly rather than simulating keystrokes, and click through a path that can fall back to invoking a click in JavaScript with no pointer movement at all.
OpenManus is rejected by Turnstile on every browser configuration we test, including the full profile: an authentic, identity-bound Chrome session is not sufficient when the control path never produces a pointer trajectory for the widget to observe.

These categories explain the Turnstile results of Table~\ref{tab:rq2_env_grid} but not the reCaptcha columns.
Invisible reCaptcha~v2 and reCaptcha~v3 are indifferent to all three control paths: every agent is rejected until the full profile is supplied, and admitted once it is, independent of how that agent types or clicks.
The two defenses therefore appear to score different properties of the session: Turnstile is sensitive to how input is generated whenever an agent's control path is not already a complete synthesis of it, while the reCaptcha family strongly relies on persistent account identity in our testbed.
Appendix~\ref{app:control-path-detail} gives the exact API calls and event traces behind each category.

Our environment ladder is instantiated on desktop Chrome, so the individual admission results should not be interpreted as browser-independent results.
However, the three properties we analyzed (browser authenticity, persistent identity, and action realization) are not inherently Chrome-specific.
They provide a framework for testing analogous effects on other browser engines, while the specific verifier dependence on each property remains an empirical question.




We next ask what remains of the defense once that environment is supplied to the agent. A trust signal is economically meaningful only if it imposes a cost that recurs on later attempts.

\section{RQ3: The Economics of Reconstructed Trust}
\label{sec:rq3_bypass}

A Captcha's security rests entirely on asymmetric cost: making automated access more expensive than the value it yields. 
The trust property used by a non-interactive defense must be sufficiently common and accessible to legitimate users to avoid imposing substantial friction. 
However, if the same property can also be acquired or reproduced by an attacker at comparable cost, its value as a security signal is limited. Our analysis therefore considers not only the cost of using the trusted environment, but also whether that trust creates recurring cost for the attacker.
The previous section established that an attacker can supply the agents with signals and properties to cross the trust boundary. 
RQ3 asks what remains of the defense once that property is supplied. Specifically, whether the defense still adds a capability cost or a per-attempt setup cost.

We restrict the economic analysis in this section to reCaptcha v3 in particular, rather than treating each symmetrically. 
Our results in the preceding sections already showed that Turnstile's trust signal can be crossed with minimal effort, leaving little recurring cost structure left to analyze. 
We therefore concentrate our experiments on reCaptcha v3, since it represents an instance of a widely deployed and purely non-interactive defense. 
The underlying arguments apply equally to reCaptcha v2 in its invisible configuration or other non-interactive defenses with a similar workflow.




\begin{table*}[t]
\centering
\small
\renewcommand{\arraystretch}{1.2}
\caption{reCaptcha~v3 outcomes for \textbf{Skyvern} and \textbf{Browser-Use} across four LLM models (100 attempts each). Every completed run cleared reCaptcha~v3 with a score of 0.9--1.0, and every failure was a submission failure rather than a score rejection, so the spread across backends reflects agent competence. Better values are in bold and relative improvements are marked.}
\label{tab:recaptcha-results}
\begin{tabular}{l|r r|r r|r r} 
\toprule
\multirow{2}{*}{\textbf{Model}} 
& \multicolumn{2}{c|}{\textbf{Completion Rate}} 
& \multicolumn{2}{c|}{\textbf{Avg Time (s)}} 
& \multicolumn{2}{c}{\textbf{Avg Used Tokens}} 
\\
\cmidrule(lr){2-3} \cmidrule(lr){4-5} \cmidrule(lr){6-7} 
& \textbf{Skyvern} & \textbf{Browser-Use} 
& \textbf{Skyvern} & \textbf{Browser-Use} 
& \textbf{Skyvern} & \textbf{Browser-Use} 
\\
\midrule
GPT-4o-mini
& \textcolor{green!60!black}{\scriptsize$\uparrow$1\%}\,\textbf{99\%} & 98\% 
& 103.42 & \textcolor{green!60!black}{\scriptsize$\downarrow$25\%}\,\textbf{77.09}  
& 77,416 & \textcolor{green!60!black}{\scriptsize$\downarrow$59\%}\,\textbf{31,857} 
\\
llama3-70b-instruct 
& 97\% & 97\% 
& 205.74 & \textcolor{green!60!black}{\scriptsize$\downarrow$34\%}\,\textbf{135.65} 
& \textcolor{green!60!black}{\scriptsize$\downarrow$75\%}\,\textbf{6,390} & 25,586 
\\
gpt-oss-120b        
& 55\% & \textcolor{green!60!black}{\scriptsize$\uparrow$78\%}\,\textbf{98\%} 
& 125.64 & \textcolor{green!60!black}{\scriptsize$\downarrow$55\%}\,\textbf{56.03} 
& \textcolor{green!60!black}{\scriptsize$\downarrow$74\%}\,\textbf{7,900} & 30,984 
\\
deepseek-v3-0324    
& \textcolor{green!60!black}{\scriptsize$\uparrow$27\%}\,\textbf{93\%} & 73\% 
& 203.69 & \textcolor{green!60!black}{\scriptsize$\downarrow$63\%}\,\textbf{75.78} 
& \textcolor{green!60!black}{\scriptsize$\downarrow$46\%}\,\textbf{6,723} & 12,468 
\\
\bottomrule
\end{tabular}
\end{table*}

\subsection{Capability Cost}
\label{sec:rq3_backend}
We hold the reconstructed environment of Section~\ref{sec:rq2_reconstruction} fixed and vary only the reasoning backend, isolating the effect of the agent from that of the defense. Changing the model affects the perception and reasoning layer, but once the reconstructed browser environment is supplied, it does not alter the verification outcome in our setting. 
The model selection directly affects completing the task and submitting the form, but not crossing the reCaptcha~v3 verification boundary. 
We evaluate four LLMs across two agent frameworks, Skyvern and Browser-Use, as the two most popular frameworks based on their GitHub repository stars and community, with 100 attempts per configuration, and report completion rate, execution time, and token usage in Table~\ref{tab:recaptcha-results}.

The results show substantial variation in workflow completion across agent--backend configurations. Skyvern achieved completion rates from 55\% to 99\%, with GPT-4o-mini reaching 99\% and llama3-70b-instruct 97\%. Browser-Use achieved comparable rates on most configurations, with deepseek-v3-0324 achieving 73\%. These differences reflect the ability of the agent and backend to complete the surrounding workflow rather than their ability to bypass reCaptcha~v3.
Across all ten agent--backend pairings, every completed login cleared reCaptcha~v3 with a score of 0.9--1.0. The lower completion rates, including 55\% for gpt-oss-120b with Skyvern and 73\% for deepseek-v3-0324 with Browser-Use, result from failures to carry the workflow through submission rather than from v3 rejection. Thus, bypass is satisfied, while completion remains LLM backend-dependent. 

Skyvern generally consumed fewer tokens but took longer, consistent with its more exploratory page interaction, whereas Browser-Use completed tasks faster while consuming more tokens. The open-source models---deepseek-v3-0324, gpt-oss-120b, and llama3-70b-instruct---were served through Novita.ai~\cite{novita_ai_2026}, whose structured-output support enables the user to map model responses to deterministic browser actions of the agent.
Overall, these results further indicate
LLM backend differences determine the efficiency and reliability of the task execution, rather than successful bypass, reflected in completion rate, execution time, and token consumption.

%

%

\subsection{Per-Attempt Setup Cost}
\label{sec:rq3_identity}
In our setup, no additional preparation cost is incurred because a grounded consumer Google account was already available and signed into authentic Chrome (Section~\ref{sec:rq2_reconstruction}).  
For an automated agent that already has such an account, incremental setup involves pointing the Chrome profile at that account and attaching CDP. 
While creating a new Gmail account is cheap, it proved insufficient for the bypass. 
Account possession raises the attacker's setup requirements, but it is not automatically a strong security boundary, because a consumer account is an attacker-accessible capability rather than a defender-controlled asset. 
Two properties keep this cost off the per-attempt ledger: the account is prepared once and reused across later attempts, and the same account is reused across the defense configurations in our testbed, with no per-configuration adaptation.
In those deployments, the trusted environment does not require a new profile for each defense. A defender cannot answer this by raising the threshold, because the agent already scores where genuine users do.

Notably, the trusted condition is not an inherently rare capability. Its usefulness for legitimate users depends on it being obtainable through ordinary account and browser use. This creates a security tension: a trust signal that is deliberately accessible to a broad population of legitimate users may also present a relatively accessible acquisition target for an adversary.
Table~\ref{tab:cost-comparison} places these findings against the two ways of bypassing reCaptcha~v3. Solver services charge a fee per challenge and, against v3, topped out at 63\% (Table~\ref{tab:solver-performance-full}). Learning-based solvers reach up to 99.6\%~\cite{tsingenopoulos2022captcha,akrout2019hacking} but pay a high fixed cost in training compute and Captcha-specific data. 
An agent supplied with a grounded profile in an authentic Chrome pays neither a per-solve wage nor a training run: the recurring cost is the task execution itself, and the one-time requirement of acquiring that account. 

\begin{table}[t]
\centering
\caption{Comparison with established reCaptcha~v3 bypass approaches. Recurring cost is incurred per attempt; setup cost is incurred once.}
\label{tab:cost-comparison}
\footnotesize
\setlength{\tabcolsep}{4pt}
\renewcommand{\arraystretch}{1.05}
\begin{tabular}{lccc}
\toprule
\textbf{Approach} & \textbf{Performance} & \textbf{Per-attempt} & \textbf{Setup} \\
\midrule
Solver services & $63\%$ & Per-solve & None \\
Learning-based & $99.6\%$~\cite{tsingenopoulos2022captcha} & Inference & Training \\
Agent + Identity & $99\%$ & Task execution & Trusted Profile\\
\bottomrule
\end{tabular}

\vspace{1mm}
\raggedright
\footnotesize
Solver-service performance from Table~\ref{tab:solver-performance} and agent rates from Table~\ref{tab:recaptcha-results}.
\end{table}

\begin{tcolorbox}[
colback=gray!10,
colframe=black!40,
title=\textbf{Key Observation},
fonttitle=\bfseries,
]

Reconstructed trust can become a reusable attacker capability.
Once a trusted profile is available to the agent, the defense does not impose a recurring cost on subsequent automated attempts. Across our experiments, the full profile transfers across agent implementations and deployments, while reCaptcha v3 contributes no measurable defense-specific overhead to workflow completion. 
Thus, the security cost of a trust-based defense can shift from a per-attempt cost to a lower acquisition cost: if the account can be reused, its cost can be amortized across many attacks. 
\end{tcolorbox}

%% file: 06-discussion.tex
\section{Discussion}

\noindent \textbf{Transferability of Trust Signals.}
Our observations suggest that the security value of a bot defense depends not only on whether its classifier can distinguish humans from automation, but also on whether the signals underlying that distinction can be reproduced and transferred by an attacker.
Challenge-based defenses expose a discrete puzzle that can be outsourced.
Identity-based defenses instead expose a trusted account or execution context that may be transferable.
This creates a tension for non-interactive defenses: the trusted condition must be sufficiently accessible to legitimate users to avoid imposing substantial friction, yet sufficiently costly for attackers to acquire or reproduce.

\noindent \textbf{Countermeasures and Defender Guidance.}
Our results motivate two groups of measures that follow from the findings rather than from isolated recommendations.

\noindent \emph{Make trust non-transferable.} Bind the verdict to a device or account that the attacker cannot move by signing the same consumer identity into Chrome, through attestation (e.g., Android Play Integrity~\cite{android_play_integrity}, Apple's DeviceCheck~\cite{apple_devicecheck}) and through checks that distinguish ordinary account possession from automated use of that account. These raise the cost of acquiring and transferring a trusted environment, not the cost of answering a challenge. Account possession is a setup requirement, not a scarce defender-controlled secret.

\noindent \emph{Treat challenge completion as weak evidence.} Token possession is not evidence of humanity: solver services return vendor-accepted tokens on challenge-based defenses at near-perfect rates. Enforce server-side, request-specific token validation so a token cannot be replayed; audit solver-service traffic; and exploit limited solver coverage as a lever against commodity pipelines when applicable.


\noindent \textbf{Security and Safety of LLM-based Agents.}
This study is a concrete example of dual-use risk in LLM agents: the goal decomposition, adaptive planning, and robust web interaction that make agents productive also lower the barrier to adversarial automation, achievable by composing public agents, solver APIs, and minor code changes rather than by training a model. Risk therefore cannot be managed at the model layer alone---browser instrumentation, action-space design, and default prompting all shape how easily an agent is repurposed---so developers should consider whether their default action spaces expose human-emulation primitives and whether their deployment environments resist fingerprinting.

\noindent \textbf{Future Directions.}
We anticipate an "AI-vs-AI" dynamic in which both attackers and defenders improve automation realism and detection quality in parallel. Future work should: (i) study how emerging browser attestation mechanisms affect the attacker cost curve by making trust non-transferable; and (ii) investigate methods to create trust boundaries for agents executed by legitimate users via a secret artifact that is only distributed by the legitimate user to an identified agent.

%% file: 07-limitations.tex
\section{Limitations}


\noindent \textbf{Adaptive and opaque scoring.}
We do not attempt to recover the providers' internal scoring function.
For instance, reCaptcha~v3 scoring is opaque and can drift over time and across sites, so an individual score cannot be attributed to a single cause.
In the studied testbed deployments, a grounded profile in authentic Chrome is a dominant transferable component of the observed trust decision.

\noindent \textbf{Model and agent scope.}
We evaluate a representative, but not exhaustive set of LLM backends and browser agents, and the landscape evolves rapidly.

\noindent \textbf{Browser generalization.}
Our browser environment analysis uses desktop Chrome and therefore does not establish the same admission pattern on Firefox, Safari, Edge, or mobile browsers.
Mobile environments further introduce distinct browser runtimes and interaction paths.
We therefore treat the Chrome results as measurements of the studied configuration, leaving cross-browser and mobile replication to future work.



%
%


%% file: 09-conclusion.tex
\section{Conclusion}
We study modern bot-management defenses from the perspective of an end-to-end automated browser workflow, asking not whether a Captcha can be solved in isolation, but whether a browser agent can reach a protected action against a live, deployed defense. This distinction matters because the two questions have different answers. Challenge-based defenses can be outsourced to commodity solver services with no reasoning capability required. On the other hand, non-interactive defenses eliminate that option and shift the bottleneck to properties of the browser session that an agent controls. We show that when the trust property is reconstructed, the same profile transfers across the agents and defense configurations in our testbed, turning the defense's decision boundary into a one-time setup cost rather than a cost that recurs on every attempt.
This leads to our central observation: the security of a non-interactive bot defense depends not only on the difficulty of reproducing its trust signal, but on whether that signal, once reproduced, is transferable. In our experiments, once a grounded profile is supplied in an authentic Chrome browser, reCaptcha~v3 imposes no additional measurable overhead beyond the underlying workflow, without a cost that recurs across attempts or targets.
The implication is that raising challenge difficulty is insufficient once the defense no longer makes its decision from a challenge at all. An agent that can complete an ordinary web workflow already possesses everything reCaptcha~v3 asks of it, provided the browser it operates in is one the defense already trusts. Defenders therefore need to shift their effort from raising the cost of solving a puzzle to raising the cost of acquiring, reproducing, and transferring the trust signal itself.

%% file: 10-appendix.tex
\section*{Appendix}

\section*{Ethical Considerations}

\noindent This work studies security weaknesses in widely deployed anti-automation mechanisms and is intended to improve defensive practice. Our experiments are designed to minimize harm by prioritizing controlled environments and avoiding disruptive behavior. We do not release exploit-ready automation scripts, sensitive endpoint details, or operational instructions that would materially lower misuse barriers.

\noindent The study follows a dual-use risk model. We report enough technical detail for scientific reproducibility and defensive validation, while withholding implementation specifics that would directly enable abuse at scale. Findings are presented as defender guidance rather than offensive playbooks.

\vfill\newpage





\section{Solver-Service Evaluation Pipeline}
\label{sec:appendix_solver_pipeline}

\noindent Section~\ref{sec:solver-service-eval-pipeline} and Figure~\ref{fig:solver_measurement} describe the measurement loop used to produce Tables~\ref{tab:solver-performance} and~\ref{tab:solver-performance-full}: one worker per service, 100 tasks per supported target, a 4-minute timeout, and vendor-side token verification as ground truth. Figure~\ref{fig:3rd_party_solver_sequence_diagram} is the complementary operational view. Captcha solutions that a service does not offer are not submitted and appear as ``--'' in the table.

\begin{table*}[t!]
\centering
\small
\setlength{\tabcolsep}{3pt}
\renewcommand{\arraystretch}{1.15}
\caption{Per-service performance of third-party solvers. Succ.\ = success rate, Time = solving time.
--: not supported. We observe a nearly perfect success rate for challenge-based defenses but only 23\% success rate on average against reCaptcha v3.}
\label{tab:solver-performance-full}
\resizebox{\textwidth}{!}{%
\begin{tabular}{l l cc cc cc cc cc cc cc cc}
\toprule
& &
\multicolumn{2}{c}{\textbf{2Captcha}} &
\multicolumn{2}{c}{\textbf{Anti-Captcha}} &
\multicolumn{2}{c}{\textbf{AZCaptcha}} &
\multicolumn{2}{c}{\textbf{BestCaptchaSolver}} &
\multicolumn{2}{c}{\textbf{CapMonster}} &
\multicolumn{2}{c}{\textbf{Capsolver}} &
\multicolumn{2}{c}{\textbf{ImageTyperz}} \\
\cmidrule(lr){3-4}\cmidrule(lr){5-6}\cmidrule(lr){7-8}\cmidrule(lr){9-10}
\cmidrule(lr){11-12}\cmidrule(lr){13-14}\cmidrule(lr){15-16}
\textbf{Captcha} & \textbf{Variant}
& \scriptsize Succ. & \scriptsize Time
& \scriptsize Succ. & \scriptsize Time
& \scriptsize Succ. & \scriptsize Time
& \scriptsize Succ. & \scriptsize Time
& \scriptsize Succ. & \scriptsize Time
& \scriptsize Succ. & \scriptsize Time
& \scriptsize Succ. & \scriptsize Time \\
& & \scriptsize (\%) & \scriptsize (s)
& \scriptsize (\%) & \scriptsize (s)
& \scriptsize (\%) & \scriptsize (s)
& \scriptsize (\%) & \scriptsize (s)
& \scriptsize (\%) & \scriptsize (s)
& \scriptsize (\%) & \scriptsize (s)
& \scriptsize (\%) & \scriptsize (s) \\
\midrule
\multirow{2}{*}{\textbf{hCaptcha}}
& Easy
  & \textbf{100} & 67.0 & -- & -- & -- & --
  & -- & -- & -- & -- & -- & --
  & \textbf{100} & 42.3 \\
& Hard
  & \textit{58} & 102.5 & -- & -- & -- & --
  & -- & -- & -- & -- & -- & --
  & \textbf{100} & 82.7 \\
\midrule
\multirow{2}{*}{\textbf{reCaptcha v2}}
& Checkbox
  & \textbf{100} & 10.8 & \textbf{100} & 51.0 & \textbf{100} & 132.2
  & \textbf{100} & 60.2 & \textbf{100} & 8.4  & \textbf{100} & 11.1
  & \textbf{100} & 49.4 \\
& Invisible
  & \textbf{100} & 13.7 & \textbf{100} & 52.7 & \textbf{100} & 101.7
  & \textbf{100} & 53.5 & \textbf{100} & 7.9  & \textbf{100} & 7.2
  & \textbf{100} & 64.3 \\
\midrule
\multirow{1}{*}{\textbf{reCaptcha v3}}
& Score ($>$0.5)
  & \textit{26} & 22.1 & \textit{30} & 12.1 & \textit{0} & 17.5
  & 63 & 39.7 & \textit{0} & 4.5 & \textit{0} & 1.6
  & \textit{41} & 40.4 \\
\midrule
\multirow{2}{*}{\makecell{\textbf{Cloudflare}\\\textbf{Turnstile}}}
& Managed
  & \textbf{100} & 7.2 & \textbf{100} & 20.3 & -- & --
  & \textbf{100} & 71.0 & \textbf{100} & 4.8 & \textbf{100} & 5.5
  & \textbf{100} & 43.3 \\
& Invisible
  & \textbf{100} & 7.8 & \textbf{100} & 26.1 & -- & --
  & \textbf{100} & 88.8 & \textbf{100} & 4.5 & \textbf{100} & 4.0
  & \textbf{100} & 46.8 \\
\bottomrule
\end{tabular}
}
\end{table*}

\begin{figure}[h]
\centering
\resizebox{\columnwidth}{!}{
\begin{tikzpicture}[
    font=\normalsize,
    box/.style={
        draw,
        align=center,
        inner sep=6pt,
        minimum width=2.6cm,
        minimum height=0.8cm
    },
    arrow/.style={
        -{Latex[length=1.6mm]},
        thick
    },
    lifeline/.style={
        dashed,
        thick
    }
]

\node[box] (web)     at (0,   0) {\textbf{Target}\\\textbf{Web Page}};
\node[box] (crawler) at (5.5, 0) {\textbf{Automated}\\\textbf{Crawler}};
\node[box] (solver)  at (11,  0) {\textbf{Third-Party}\\\textbf{Solver}};

\draw[lifeline] (web.south)     -- ++(0,-9.5);
\draw[lifeline] (crawler.south) -- ++(0,-9.5);
\draw[lifeline] (solver.south)  -- ++(0,-9.5);

\draw[arrow] ([yshift=-1.0cm]crawler.south)
    -- node[above,font=\small]
       {\textbf{\textcircled{\raisebox{-0.9pt}{1}}}~~Visit Page}
    ([yshift=-1.0cm]web.south);

\draw[arrow] ([yshift=-2.5cm]web.south)
    -- node[below,font=\small,align=left]
       {\textbf{\textcircled{\raisebox{-0.9pt}{2}}}~~Get Parameters\\
        \hspace*{1.4em}$-$ SiteKey\\
        \hspace*{1.4em}$-$ Captcha Type}
    ([yshift=-2.5cm]crawler.south);

\draw[arrow] ([yshift=-4.8cm]crawler.south)
    -- node[above,font=\small,align=left]
       {\textbf{\textcircled{\raisebox{-0.9pt}{3}}}~~Create Task\\
        \hspace*{1.4em}$-$ SiteKey\\
        \hspace*{1.4em}$-$ Captcha Type\\
        \hspace*{1.4em}$-$ Service Parameters}
    ([yshift=-4.8cm]solver.south);

\draw[arrow] ([yshift=-6.6cm]solver.south)
    -- node[above,font=\small,align=left]
       {\textbf{\textcircled{\raisebox{-0.9pt}{4}}}~~Send Solution\\
        \hspace*{1.4em}$-$ Token}
    ([yshift=-6.6cm]crawler.south);

\draw[arrow] ([yshift=-7.8cm]crawler.south)
    -- node[above,font=\small,align=left]
       {\textbf{\textcircled{\raisebox{-0.9pt}{5}}}~~Use the solution\\
        \hspace*{1.4em}$-$ Form Values\\
        \hspace*{1.4em}$-$ Token}
    ([yshift=-7.8cm]web.south);

\draw[arrow] ([yshift=-9.2cm]web.south)
    -- node[above,font=\small]
       {\textbf{\textcircled{\raisebox{-0.9pt}{6}}}~~Send Results:
        Success or Failure}
    ([yshift=-9.2cm]crawler.south);

\end{tikzpicture}
}
\caption{Sequence diagram of a third-party Captcha solver integrated into an automated crawler. The crawler extracts challenge parameters, delegates solving to the service, and injects the returned token before submitting the target form.}
\label{fig:3rd_party_solver_sequence_diagram}
\end{figure}


\section{Testbed Login Pages}

\label{sec:appendix_testbed_pages}
\begin{figure}[h]
    \centering
    \includegraphics[width=0.9\linewidth]{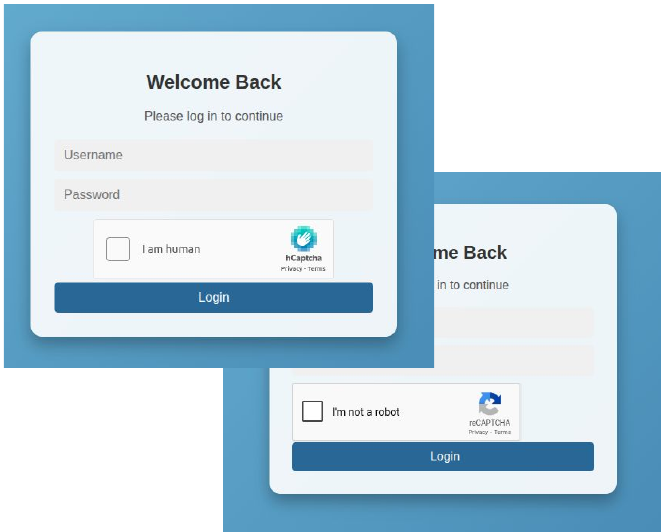}
    \caption{Sample login pages from our controlled testbed, protected with hCaptcha and reCaptcha~v2.}
    \label{fig:sample_login_v2_hCaptcha}
\end{figure}

\section{Prevalence Crawl}
\label{sec:appendix_prevalence}

\noindent \textbf{Sampling and scope.}
We drew our candidate sites from the Chrome User Experience Report (CrUX) dataset~\cite{google_crux}, which ranks origins by real-user popularity, and crawled the \emph{landing page} of each. Of the origins we attempted, 350{,}174 landing pages were retrieved successfully; the remainder failed to resolve, timed out, or returned a non-HTML response, and are excluded. The crawl was conducted based on the February 2026 CrUX release, over the top 500k websites.
Restricting the crawl to landing pages is a deliberate scope limit with a known direction of bias. Bot-management widgets cluster on pages carrying forms, sign-up, login, password reset, and contact, and many sites expose none of these on the landing page itself. Enumerating every form-bearing page across 350{,}174 sites would require a crawling infrastructure well beyond this study. Our counts should therefore be read as a lower bound on deployment, and the ratios between vendors rather than the absolute totals are the quantity we rely on.

\noindent \textbf{Detection.}
We identify a deployment by the provider scripts and widget parameters embedded in the returned HTML, which distinguishes the deployed defense service. Detection is therefore a statement about what a page loads, not about whether the site enforces the resulting verdict. Because the signature is the provider's own script, we expect false positives to be rare; false negatives are more likely, since a site that injects its widget after user interaction, or behind a consent gate, will not be counted. We did not construct a labeled ground-truth set, so we do not report a validated detection rate, and the counts below should be read with that caveat.


\noindent \textbf{Results.}
Of the 350{,}174 retrieved landing pages, 35{,}786 embed at least one of the four services. reCaptcha~v3 accounts for 24{,}073 sites, over two-thirds (67\%) of all detected deployments, ahead of reCaptcha~v2 (8{,}908; 25\%), Cloudflare Turnstile (1{,}642; 5\%), and hCaptcha (1{,}164; 3\%).

\section{Agent Task Prompt}
\label{sec:appendix_prompt}
\begin{figure}[h]
  \centering
  \begin{tcolorbox}[listing only,
    colback=promptbg,
    colframe=promptborder,
    listing options={
      basicstyle=\ttfamily\small,
      breaklines=true,
      columns=fullflexible,
      language=,
      escapeinside=||,
    },
    left=0.5mm,
    right=0.5mm,
    top=0.5mm,
    bottom=0.5mm,
    boxrule=0.8pt,
    arc=2pt]
Navigate to https://example-login.com and fill out the form with the following information: \\
-----\\
username: [username], password: [password] \\
-----\\
If you encounter a verification challenge, attempt to solve it before submitting. \\
Submit the form by clicking the login button.
  \end{tcolorbox}
  \caption{\normalsize
{Prompt used to instruct the LLM-based browser agents during the experiment. The URL, username, and password have been replaced with placeholders.}}
  \label{fig:cracking_prompt}
\end{figure}

\begin{figure}[h]
    \centering
    \includegraphics[width=\linewidth]{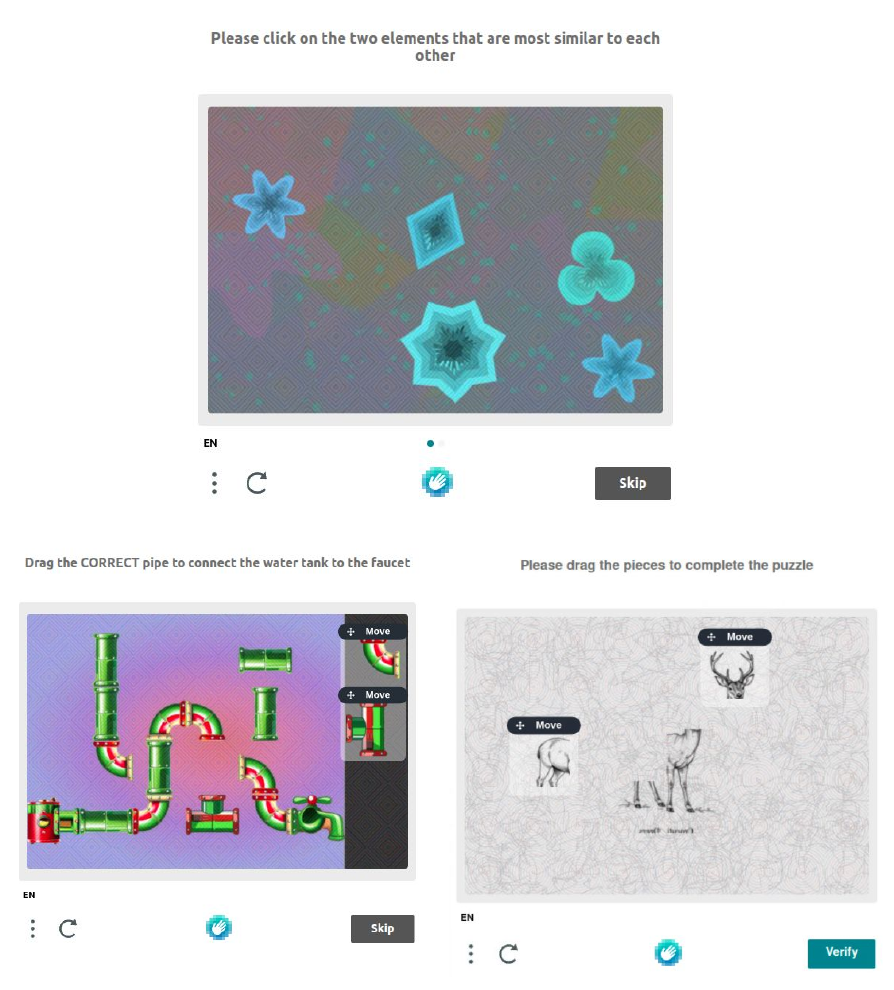}
    \caption{Samples of the visual puzzles served during hCaptcha evaluation: a similarity-selection task (top), a pipe-routing drag puzzle (bottom left), and a piece-placement drag puzzle (bottom right). No agent cleared these. In the architecture of Figure~\ref{fig:llm_browser_agent_pipeline} these are mainly perception and reasoning failures: the challenge is rendered and the agent engages it, but the model does not land the required selection or placement.
    }
    \label{fig:hcaptcha_samples}
\end{figure}


\section{Control-Path Implementation Detail}
\label{app:control-path-detail}

This appendix supplements Section~\ref{sec:recaptcha_behavior_vs_env}'s three control-path categories with the specific mechanisms each agent uses to turn a planned action into browser-visible input.

\noindent\textbf{Browser-Use.}
Browser-Use drives the page through a CDP action watchdog that calls \texttt{dispatchKeyEvent} and \texttt{dispatchMouseEvent} directly. Page traces show a \texttt{keydown}/\texttt{keypress}/\texttt{keyup} run for every character typed and pointer events preceding every click, whether the agent is attached to its bundled Chromium or to genuine Chrome.

\noindent\textbf{Skyvern.}
Skyvern also injects keystroke-like events, but through Playwright rather than raw CDP: \texttt{locator.type} on the default text-entry path and \texttt{locator.click} for clicks. This is a different input transport from Browser-Use's CDP watchdog even though both ultimately cause the page to observe typed input and pointer events.

\noindent\textbf{OpenManus.}
OpenManus wraps an older Browser-Use \texttt{input\_text} tool that, for ordinary \texttt{input} fields, calls Playwright's \texttt{fill()}: the value is assigned in one shot, and the page sees \texttt{input}/\texttt{change} events with little or no keystroke sequence. Its clicks use \texttt{element.click()} and can fall back to a JavaScript-invoked \texttt{el.click()}, which produces a \texttt{click} event with no preceding \texttt{mousemove}/\texttt{mousedown}/\texttt{mouseup} sequence.

\noindent\textbf{SeeAct.}
SeeAct's perception is screenshot-based (set-of-marks over Playwright-enumerated elements), but its actuation is not a click at a predicted coordinate: the chosen locator is driven with \texttt{selector.click()} and \texttt{selector.fill()}. The click path matches Skyvern's framework-level synthesis; the fill path matches OpenManus's one-shot value assignment. SeeAct therefore straddles two of the three categories in Section~\ref{sec:recaptcha_behavior_vs_env}, which is consistent with it patterning with Skyvern under Turnstile (click-driven) while sharing OpenManus's lack of a pointer path on reCaptcha~v2, where it remains rejected even on the full profile.

\noindent\textbf{Why the categories predict Turnstile but not reCaptcha.}
Both agents in the instrumented configuration run a bundled Chromium rather than \texttt{google-chrome}, so Browser-Use's admission in that cell cannot be attributed to a real Chrome binary. What differs is that Browser-Use remains on the CDP \texttt{Input} domain in every configuration, while instrumented Skyvern drives stock Playwright Chromium through locator APIs; only the latter is refused. This isolates the locus of the Turnstile signal to the input transport itself, not merely to the presence of a genuine Chrome binary. reCaptcha~v2 and reCaptcha~v3, by contrast, reject every agent in every category until the full profile is supplied and admit all of them once it is (Table~\ref{tab:rq2_env_grid}), which is why the categorization drawn from Turnstile does not carry over to the reCaptcha rows.